\documentclass{optica-article}

\journal{opticajournal} 

\articletype{Research Article}

\usepackage{lineno}
\usepackage{chemformula} 
\usepackage[T1]{fontenc} 
\usepackage{siunitx}
\usepackage{booktabs}
\usepackage{multirow}
\usepackage{subfig}
\usepackage[]{graphicx}
\usepackage{MnSymbol,wasysym}
\usepackage{soul,xcolor}

\begin{document}

\title{Combining Bayesian Optimization, SVD and Machine Learning for Advanced Optical Design}


\author{M. R. Mahani\authormark{1,*}, Igor A. Nechepurenko\authormark{1}, Thomas Flisgen\authormark{1,2} , Andreas Wicht\authormark{1}}

\address{\authormark{1}Ferdinand-Braun-Institut (FBH), Gustav-Kirchhoff-Straße 4, 12489, Berlin, Germany}

\address{\authormark{2}Brandenburgische Technische Universität Cottbus - Senftenberg, Fachgebiet Theoretische Elektrotechnik, Siemens-Halske-Ring 14, 03046 Cottbus, Germany}

\email{\authormark{*}Reza.Mahani@FBH-Berlin.de}

\begin{abstract*} 
The design and optimization of optical components, such as Bragg gratings, are critical for applications in telecommunications, sensing, and photonic circuits. To overcome the limitations of traditional design methods that rely heavily on computationally intensive simulations and large datasets, we propose an integrated methodology that significantly reduces these burdens while maintaining high accuracy in predicting optical response. First, we employ a Bayesian optimization technique to strategically select a limited yet informative number of simulation points from the design space, ensuring that each contributes maximally to the model's performance. Second, we utilize singular value decomposition to effectively parametrize the entire reflectance spectra into a reduced set of coefficients, allowing us to capture all significant spectral features without losing crucial information. Finally, we apply XGBoost, a robust machine learning algorithm, to predict the entire reflectance spectra from the reduced dataset. The combination of Bayesian optimization for data selection, SVD for full-spectrum fitting, and XGBoost for predictive modeling provides a powerful, generalizable framework for the design of optical components.
\end{abstract*}

\section{Introduction}
The design and optimization of optical components, such as Bragg gratings \cite{kashyap2009fiber}, play a crucial role in a wide range of applications, including telecommunications \cite{hill1997fiber,bazakutsa2023highly}, sensing \cite{cusano2011fiber,butov2022tilted}, and photonic circuits \cite{joannopoulos1997photonic,burla2013integrated}. Bragg gratings, which are periodic structures used to reflect specific wavelengths of light, require precise control over their optical properties to achieve desired performance \cite{erdogan1997fiber}. Mainstream methods for designing these components often involve extensive computational simulations and large datasets, which can be time-consuming and resource-intensive \cite{taflove2005computational,gnan2006modelling,malkiel2018plasmonic, tahersima2019deep}. As the complexity and dimensionality of the design space increase, more efficient approaches are required to balance accuracy with computational efficiency. To address the challenges inherent in conventional approaches, our study combines the advances on three key aspects of the optical design process using machine learning (ML), namely, predictive modeling, data selection, and spectral fitting.
 
ML and optimization techniques for design have gained considerable attention for their ability to model and predict complex photonics systems \cite{ahn2022photonic, white2022inverse, liu2021tackling, ma2021deep, wiecha2021deep}. Shallow ML algorithms can struggle with capturing intricate dependencies in very high-dimensional design spaces \cite{scholkopf2002learning, gunn1998support, cortes1995support}. On the other hand, deep learning methods, while powerful, often require large datasets and significant computational resources \cite{nadell2019deep, ma2021deep, hammond2019designing, hegde2020deep}, making them computationally inefficient when the data has to be acquired through simulations or experiments.
Among these ML algorithms, optimized extreme Gradient Boosting (XGBoost) \cite{chen2016xgboost} has emerged as a leading algorithm to efficiently handle both linear and nonlinear relationships while trained on limited data. XGBoost has successfully outperformed various ML models in domains with scarce data \cite{zou2022optimized}. 

Apart from a data-efficient ML algorithm, one of the crucial requirements for an accurate prediction, is the data used for ML training. For data generation in optical design, full-wave electromagnetic simulations can be used to model the reflectance spectra of Bragg gratings\cite{mahani2023data,nechepurenko2023finite}. While accurate, these methods can be prohibitively expensive, particularly when exploring large parameter spaces or optimizing multiple design variables simultaneously. Large databases (of the order of $10^5$) for training deep learning models are widely used and to generate these large databases, faster and less accurate simulations are used \cite{hammond2019designing}. However, we have previously proposed a novel approach, based on Bayesian optimization, that significantly reduces the number of required simulations for training ML models while maximizing model performance \cite{mahani2024optimizing}. It was shown that the same ML performance can be obtained, using an order of magnitude fewer data for training and this could further improve in more complex design space.

The output of ML algorithms is the final area that we aim to transform. In systems that we are after, their optical response is the output that the ML model is trained on, and which it predicts. Generally the spectra are either discretized to several data points at various frequencies or parameterized using a fit function, then predicted using a regression model. The former approach requires large number of data points at various frequencies otherwise the predicted spectra do not overlap with the true spectra. In the latter approach one selects the most relevant part of the spectra (top half of the main lobe, or main lobe and side lobes) and use appropriate functions (e.g., Gaussian function or coupled mode theory) to fit the selected part and extract the fit parameters. These parameters can then be used as output for training and predicting by regression algorithms \cite{mahani2023data, mahani2024tailor}. Here, we use singular value decomposition (SVD) to be able to predict the entire spectrum, rather than a small part of it. SVD is a powerful technique for dimensionality reduction, which can transform high-dimensional data into a set of orthogonal components, or singular vectors, that capture the most significant features of the data \cite{gustavsen1999rational, markovsky2007overview}. We take advantage of this method in optical design, where the spectra can be complex and multi-peaked, and conventional fitting techniques fail to capture all relevant features accurately. SVD allows for the effective parametrization of the spectra into a limited set of coefficients, significantly reducing the complexity of the data without losing essential information.

Combining all areas above, we first apply a Bayesian method \cite{garnett2023bayesian, shahriari2015taking} to strategically select data points, ensuring that each contributes optimally to the model training, in contrast to conventional approaches that rely on dense, exhaustive datasets. Second, we use SVD to fit the entire reflectance spectrum, unlike methods such as Gaussian fitting or coupled mode theory that capture only specific features or portions of the spectrum \cite{mahani2023data, mahani2023designing}. Finally, we incorporate XGBoost, a powerful ML algorithm, to accurately predict the reflectance spectra using the reduced, highly informative dataset. This combination of advanced data selection, full-spectrum fitting, and robust ML prediction enables highly accurate predictions of the entire spectra, even with datasets as small as $10^2-10^3$. Through these improvements, we demonstrate a powerful and efficient methodology for the design and optimization of optical components that can be applied to a wide range of problems. 

\section{Data Generation}
\begin{figure}[htbp]
\centering
\includegraphics[height=1.25 in]{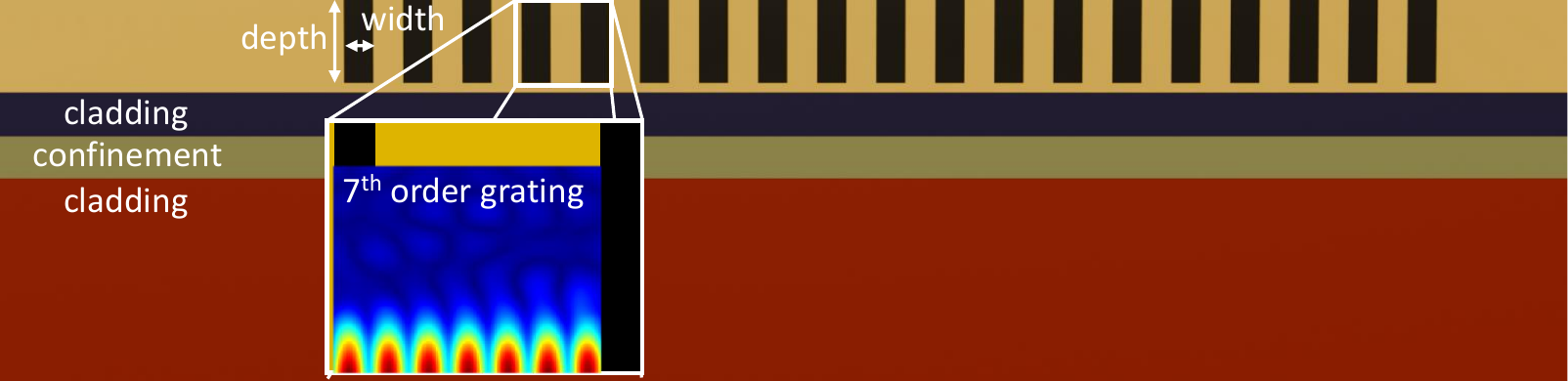}
\caption{Schematic of the structure. The inset shows the distribution of electric field magnitude for the $7^{th}$ order Bragg grating.}
\label{fig:struct}
\end{figure}

We consider a 2D GaAs ridge waveguide designed for single-mode operations. The waveguide includes $7^{th}$, $10^{th}$, $13^{th}$ and $16^{th}$ order Bragg grating. The order number refers to the mode of diffraction, reflection, for which the Bragg condition is satisfied. They are composed of periodic rectangular grooves, with a total length of 200 $\mu$m. The Bragg grating width coincides with the ridge width, while the duty cycle and groove depth are variable parameters influencing the grating's reflectivity, which peaks at a frequency of $280$ THz, close to a wavelength of $1064$ nm. The schematic of the structure with relevant parameters is depicted in figure \ref{fig:struct}. The details of the simulation are explained at length in Refs. \cite{mahani2023data, rahimof2024study, nechepurenko2023finite}

In order to build a database of many simulations with varied parameters, Lumerical and Matlab integration was utilized to perform 2736 ($4 \times 76 \times 9$) two-dimensional (2D) finite-difference time-domain (FDTD) simulations. Each simulation was done with a mesh size of 20 nm and a time step of 0.05 fs.
The simulation parameters were chosen to ensure fabrication feasibility, focusing on the groove depth and width and order number as key variables affecting the reflectance and coupling coefficient. 
As mentioned above, we choose four different order numbers, the depth was sampled with $76$ linearly spaced numbers in the interval $1.08$ $\mu$m to $1.42$ $\mu$m, and the width was sampled with $9$ linearly spaced values in the interval $20$ nm to $100$ nm. We explored the design space by generating 2736 distinct structures (obtained by these variations) and ran simulations to calculate the corresponding reflectance spectra. The design parameters were used as inputs, while the resulting reflectance spectra served as outputs (with further post processing explained below), which were compiled into a database for further analysis. These simulations distributed uniformly on each parameter set were generated to be used by Bayesian selection, explained in section \ref{bayesian}. As we explain later we only need a fraction of these simulations (chosen by Bayesian selection) to train ML.

\section{Singular Value Decomposition Fit}
In order to be able to predict the reflectance spectra (that normally contain many data points), one can fit the spectra using various functions.
This would allow us to parameterize the reflectance spectra (with a certain accuracy) and reconstruct it with few parameters (fit coefficients). We have previously reported fitting the top two-third of the main lobe of spectra using a Gaussian function \cite{mahani2023data}. This would give us three fit parameters to reconstruct the top of the main lobe. We then train a ML algorithm on the database built by the design parameters as input and fit parameters as output (three in this case) \cite{mahani2023data, mahani2023designing}. In order to be able to predict a larger section of the spectra, we have also used coupled-mode-theory (CMT) \cite{mahani2024tailor, rahimof2024coupled}. One could predict the main-lobe and the two-side lobes of the spectra using CMT with a reasonable accuracy, and then train a ML model on the three CMT fit parameters.

However thus far, to the best of our knowledge, the prediction of the full spectra with a great accuracy has not been reported. Here we aim to fit the entire spectra using SVD method, well-known to the research communities who deal with signal processing \cite{moonen1995svd, le2004singular}. SVD is a powerful linear algebra technique that decomposes a matrix into three constituent matrices, capturing the essential patterns in the data while reducing noise and redundancy \cite{gustavsen1999rational, markovsky2007overview, golub1971singular}. We employed SVD to fit the optical reflectance spectra obtained from the simulations explained in previous section. The details of the fit is explained below.

\subsection{Reflectance Data Matrix}
Let $\mathbf{R} \in \mathbb{C}^{M \times N}$ be the reflectance data matrix, where each row corresponds to a different reflectance spectrum (by varying the design parameters), and each column represents the reflectance at a specific frequency:
\begin{equation}
\mathbf{R} = \begin{pmatrix}
r_{11} & r_{12} & \cdots & r_{1N} \\
r_{21} & r_{22} & \cdots & r_{2N} \\
\vdots & \vdots & \ddots & \vdots \\
r_{M1} & r_{M2} & \cdots & r_{MN}
\end{pmatrix}\textnormal{.}\label{eq:reflectance_matrix}
\end{equation}
In our case, we have $M = 2736$ spectra and $N = 300$ number of frequency points.

\subsection{Singular Value Decomposition}
The SVD of the complex-valued matrix (\ref{eq:reflectance_matrix}) is given by
\begin{equation}
\mathbf{R} = \mathbf{U} \mathbf{S} {\mathbf{V}}^\mathrm{H}\textnormal{,}
\end{equation}
where:
\begin{itemize}
    \item $\mathbf{U}\in \mathbb{C}^{M \times M}$ is a square complex-valued unitary matrix,
    \item $\mathbf{S}\in \mathbb{R}^{M \times N}$ is a rectangular diagonal matrix containing the non-negative and real-valued singular values, sorted in descending order,
    \item $\mathbf{V}\in \mathbb{C}^{N \times N}$ is a square complex-valued unitary matrix.
\end{itemize}
Here, ${\mathbf{V}}^\mathrm{H}$ denotes the conjugate transpose of ${\mathbf{V}}$. The matrix $\mathbf{U}$ contains the weighting factors of the basis vectors and the columns of $\mathbf{V}$ represent the basis vectors used to reconstruct the spectra in the frequency domain. The diagonal entries of $\mathbf{S}$ are the singular values that quantify the significance of each basis vector.

\subsection{Truncation and Approximation}
To have a small number of degrees of freedom to describe the $M$ spectra, we truncate the SVD to retain only the first $K$ singular values, where $K$ represents the number of degrees of freedom (DOFs) chosen based on the desired accuracy of the fit. The truncated SVD approximation is given by:
\begin{equation}
\mathbf{R}_{\mathrm{approx}} = \mathbf{U}_K \mathbf{S}_K {\mathbf{V}}_K^\mathrm{H}\textnormal{,}
\end{equation}
where:
\begin{itemize}
    \item $\mathbf{U}_K\in \mathbb{C}^{M \times K}$ contains the first $K$ columns of $\mathbf{U}$,
    \item $\mathbf{S}_K\in \mathbb{R}^{K \times K}$ contains the first $K$ columns and rows of $\mathbf{S}$, and
    \item $\mathbf{V}_K\in \mathbb{C}^{N \times K}$ contains the first $K$ columns of $\mathbf{U}$.
\end{itemize}

In our study, we selected $K = 7$, retaining the most significant modes while discarding higher-order components that primarily represent less prominent features of the spectra. This number of DOFs provides a balance between the accuracy of the fit and manageable number of parameters. In total we have $2 K = 14$ parameters for the ML training and prediction, since each DOFs is described by its real and imaginary part. Lower number of DOFs has reduced accuracy and higher number of DOFs requires more data for training the ML.

The reflectance spectrum for each structure can be reconstructed using the truncated SVD as follows:
\begin{equation}
\mathbf{R}_{\text{approx},i} = \sum_{j=1}^{K} U_{ij} \mathbf{W}_j\textnormal{.}\label{eq:superposition}
\end{equation}
We refer to
\begin{equation}
\mathbf{W}_j = S_{jj}\mathbf{V}_j^\mathrm{H},
\end{equation}
as the $j$-th SVD mode, $\mathbf{R}_{\text{approx},i}\in\mathbb{C}^{1 \times N}$ represents the $i$-th row of the reconstructed reflectance matrix, i.e.\ the $i$-th reflectance spectrum. In the equation $U_{ij}$ is the element $ij$ of the matrix $\mathbf{U}$, $S_{jj}$ is the $j$-th singular value, and $\mathbf{V}_j^\mathrm{H}$ is the $j$-th row of $\mathbf{V}^\mathrm{H}$. Following (\ref{eq:superposition}), $U_{ij}$ can be considered as the complex-valued weighting coefficient of the $j$-th basis vector to reconstruct the $i$-th spectrum.

\subsection{Practical Considerations}
The matrices $\mathbf{S}$ and $\mathbf{V}$ and thus $\mathbf{W}_j$ capture the shared characteristics of the entire dataset and are thus not part of the training or prediction process of the ML. However, they are crucial for reconstructing the reflectance spectra after the weighting coefficients $U_{ij}$ are predicted by the ML model. By utilizing SVD, we can efficiently represent the complex reflectance spectra with a reduced set of parameters, more suitable for ML training yet maintaining high accuracy in predicting the entire spectra.

\begin{figure}[htbp]
\centering
\includegraphics[height=2.in]{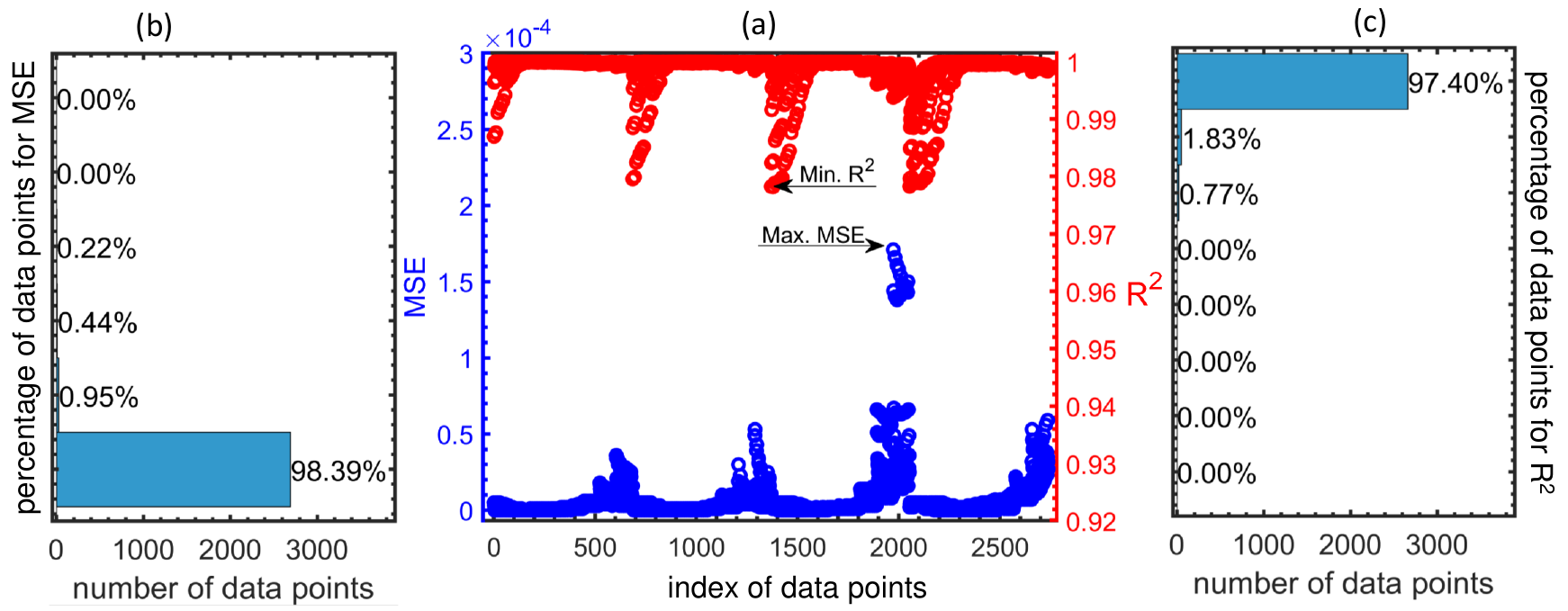}
\caption{Fit metrics and statistics. (a) This panel shows the overall performance of the SVD-based fitting model, where both MSE and R\textsuperscript{2} are visualized simultaneously, highlighting the maximum error and minimum explained variance. MSE is plotted in blue circles with its axis on the left and $R^2$ is plotted in red circles with its axis on the right. (b) This panel demonstrates the error distribution, where the majority of MSE values cluster in the lower interval, indicating that the model generally has low errors. The percentage of data points whose fit has an MSE in the interval [0, $5 \times 10^{-5}$) is 98.39\% which includes most of the data. (c)  This panel shows the R\textsuperscript{2} distribution, with most values concentrated near 1, which implies that the model explains the majority of the variance in the data, with limited number of outliers having lower R\textsuperscript{2} values. The percentage of data points whose fit has an $R^2$ in the interval [0.99, 1) is 97.40\%, which includes most of the data.}
\label{fig:fit_stats}
\end{figure}

\begin{figure}[htbp]
\centering
\includegraphics[height=3.3 in]{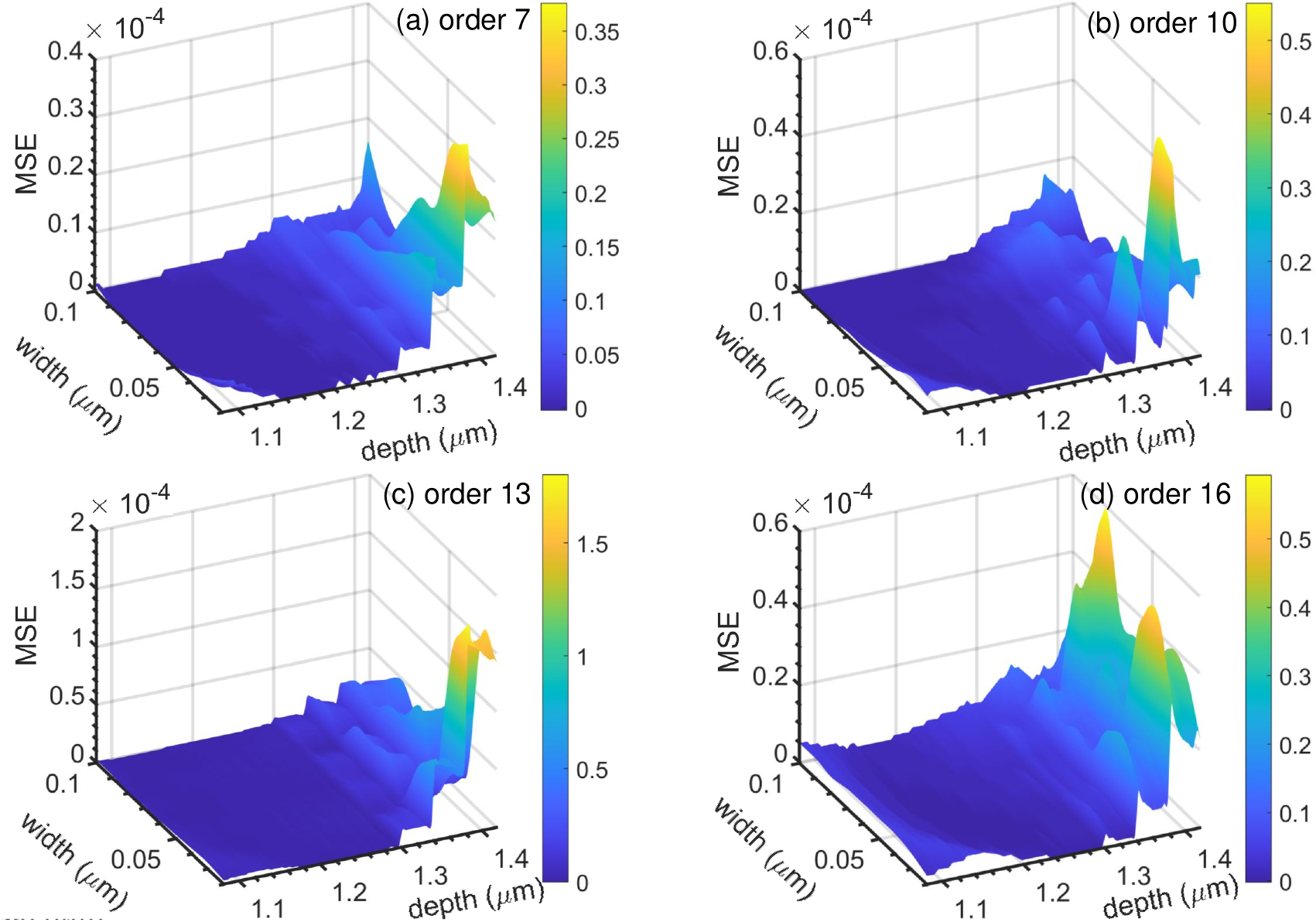}
\caption{Fit error (MSE) as a function of geometric design parameters. (a)-(d) Surface plot showing the MSE as a function of width and depth for order numbers 7, 10, 13 and 16, respectively. The z-axis represents MSE values, a factor of $10^{-4}$ appears on top of the axis; thus an MSE value of 0.2 on the axis corresponds to $2 \times 10^{-5}$.}
\label{fig:fit_stats2}
\end{figure}

We analyzed the fit accuracy of the entire dataset using the  mean squared error (MSE) and R\textsuperscript{2} (coefficient of determination) values as key metrics and provide the statistics in Figures \ref{fig:fit_stats} and \ref{fig:fit_stats2}. The analyses involve the statistics and the dependency of the fit on various input parameters. These metrics are defined as,
\begin{equation}
R^2 = 1 - \frac{\sum_{i=1}^{n} (y_i - \hat{y}_i)^2}{\sum_{i=1}^{n} (y_i - \bar{y})^2},
\end{equation}
\begin{equation}
\text{MSE} = \frac{1}{n} \sum_{i=1}^{n} (y_i - \hat{y}_i)^2\textnormal{,}
\end{equation}
where
\begin{itemize}
\item 
$y_i$ are the observed values,
\item 
$\hat{y}_i$ are the predicted values,
\item 
$\bar{y}$ is the mean of the observed values,
\item 
$n$ is the number of observations.
\end{itemize}

Figure \ref{fig:fit_stats}(a) plots the MSE and R\textsuperscript{2} values for the selected dataset, showing both metrics in a single, combined figure. The MSE represents the average squared difference between the actual and fitted spectra, while R\textsuperscript{2} quantifies the proportion of variance in the data explained by the fitted model.
In this figure MSE is plotted on the left vertical axis, in blue, to indicate the magnitude of the error in the fit. R\textsuperscript{2} is plotted on the right  vertical axis, in red, showing how well the model explains the variation in the reflectance spectra.
Additionally, key points, such as the maximum MSE and minimum R\textsuperscript{2}, are highlighted with labels and arrows in the figure, providing visual clue to the data points with the worst metrics.

The figure \ref{fig:fit_stats}(b) provides a statistical distribution of the MSE values. Here, the MSE values were divided into intervals, each representing a specific range of error values. The intervals are defined as:
\[
[0, 5\times 10^{-5}), [5\times 10^{-5}, 1\times 10^{-4}), ... , [2.5\times 10^{-4}, 3\times 10^{-4}).
\]
The number of data points with MSE values in each interval was then counted and plotted as a horizontal bar graph (total number of data points is 2736). For each interval, the percentage of data points in that interval is also calculated and displayed next to each bar. This figure provides a statistical distribution of the error in the fit, showing that the model tends to perform very well across the dataset. The percentage of data points whose fit has an MSE value smaller than $5 \times 10^{-5}$ is more than 98\% of the entire data. The MSE value for the worst fit is $1.71 \times 10^{-4}$, and the intervals with the MSE values beyond $2 \times 10^{-4}$ show zero percentage of the data.

Figure \ref{fig:fit_stats}(c) depicts the distribution of the R\textsuperscript{2} values, similar to the MSE analysis. The R\textsuperscript{2} values were divided into intervals from 0.92 to 1.0, in steps of 0.01:
\[
[0.92, 0.93), [0.93, 0.94), [0.94, 0.95), \dots, [0.99, 1.0).
\]
For each interval, the number of data points was counted and plotted as a horizontal bar graph. Like the MSE plot, the percentage of data points in each interval is displayed next to the corresponding bar, allowing for easy comparison across the intervals. The percentage of data points whose fit has an $R^2$ value above 0.99 is more than 97\% of the entire data. The worst $R^2$ value is 0.978 and the intervals with values smaller than 0.97 show zero percentage.

Figure \ref{fig:fit_stats2} is interpolated 3D surface plots representing the relationship between MSE with depth, and width for various order numbers. Figure \ref{fig:fit_stats2}(a)-(d) corresponds to order $7$, $10$, $13$ and $16$, respectively. The vertical axis shows the MSE values, while the depth and width are represented on the horizontal axes, respectively.

From the Figure \ref{fig:fit_stats2}, one can observe the continuous variation in MSE as the depth and width change, providing a smooth surface that shows regions where the error is minimized or maximized. The highest MSE values are generally observed at the extremities, suggesting that the maximum values of the depth lead to more complicated spectra and higher error for various order number.

To visualize the fit, we select a few cases. When selecting the worst fits for visualization, we focus mostly on high MSE values which provides an understanding of where the model is underperforming. Metric $R^2$ tells us how well the model explains the variance in the data, and the fits are mostly distributed in very high $R^2$ values. On the other hand MSE directly measures the average squared error between the fit and actual values, making it a reliable indicator of large deviations in the fit. Thus, MSE is more relevant when identifying specific problematic cases.
\begin{figure}[htbp]
\centering


\includegraphics[height=5.8in]{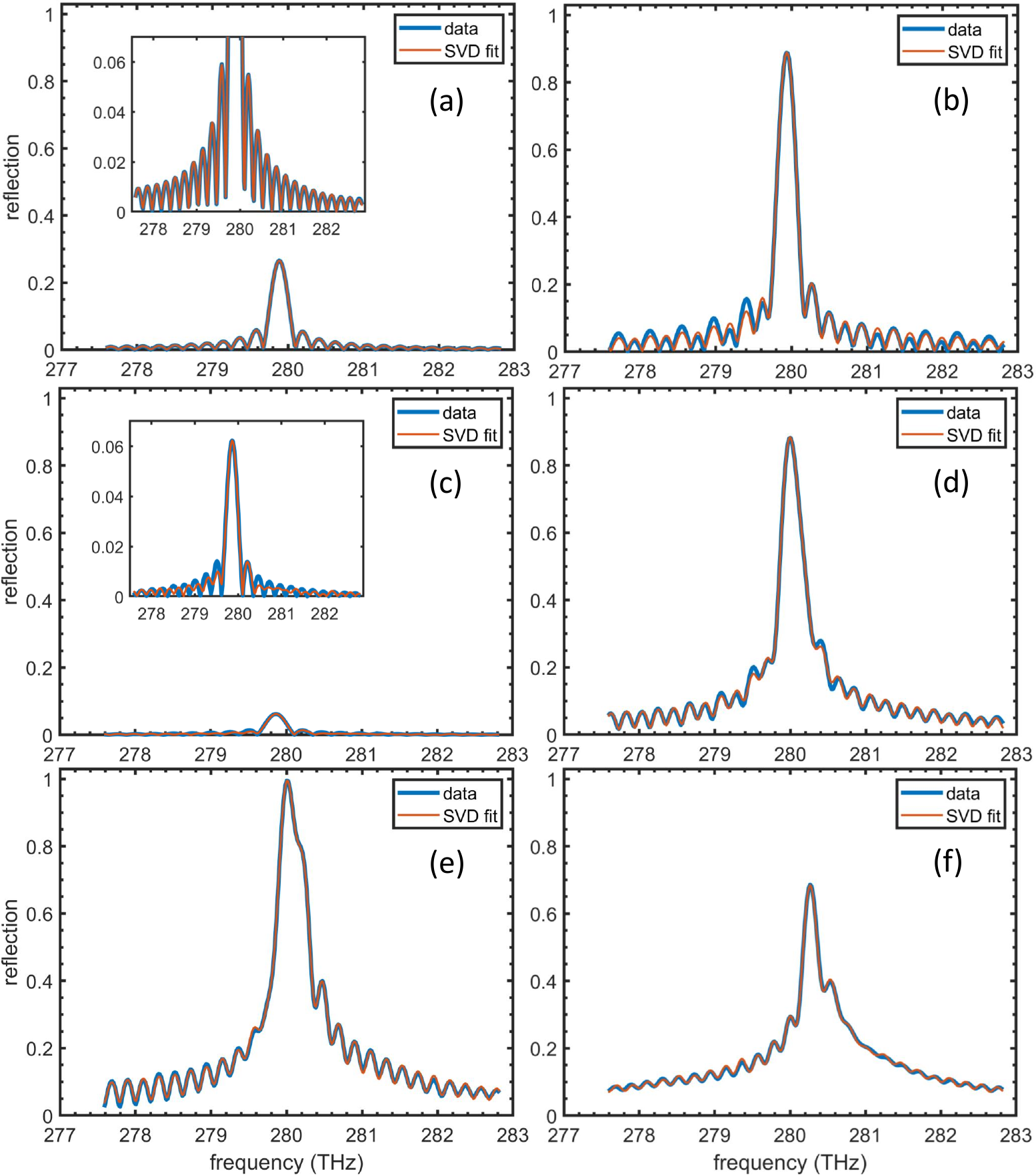}
\caption{Fit comparison for some specific cases. Solid blue line is the original spectra and solid red line is the fitted spectra using the truncated SVD. (a), (b) The data points with the best MSE ($\textnormal{MSE}=4.45 \times 10^{-8}$, $R^2=0.99998$), the inset is the zoomed-in version, and the worst MSE values for the fit ($\textnormal{MSE}=1.71 \times 10^{-4}$, $R^2=0.994$), respectively. (c) The data point with the worst $R^2$ value ($\textnormal{MSE}=3.06 \times 10^{-6}$, $R^2=0.978$), the inset is the zoomed-in version. (d) The data point with the worst MSE value taken from the first interval, $[0, 5\times 10^{-5})$, ($\textnormal{MSE}=4.95 \times 10^{-5}$, $R^2=0.998$). (e), (f) Two randomly chosen data points with the metrics ($\textnormal{MSE}=2.17 \times 10^{-5}$, $R^2=0.9995$) and ($\textnormal{MSE}=1.6 \times 10^{-5}$, $R^2=0.9989$), respectively.}
\label{fig:ref_pred}
\end{figure} 

In Figure \ref{fig:ref_pred} we plotted six spectra to visualize the accuracy of the fit. Figures \ref{fig:ref_pred}(a), (b) show the data points for which the SVD fit to the reflectance spectra has the smallest MSE ($\textnormal{MSE}=4.45 \times 10^{-8}$, $R^2=0.99998$), and the largest MSE values ($\textnormal{MSE}=1.71 \times 10^{-4}$, $R^2=0.994$), respectively. Figure \ref{fig:ref_pred}(c) shows the data point with the worst $R^2$ value ($\textnormal{MSE}=3.06 \times 10^{-6}$, $R^2=0.978$). 
Although there is a visible difference between the fit and the original data in this panel, the reflection is very small (maximum reflection around 0.06).
Figure \ref{fig:ref_pred}(d) shows the data point with the worst MSE value taken from the first interval, $[0, 5\times 10^{-5})$, ($\textnormal{MSE}=4.95 \times 10^{-5}$, $R^2=0.998$). 
Since more than 98\% of the data points are in this interval, this visually shows the accuracy of most of the fits.
Figures \ref{fig:ref_pred}(e) and (f) show two randomly chosen data points with the metrics ($\textnormal{MSE}=2.17 \times 10^{-5}$, $R^2=0.9995$) and ($\textnormal{MSE}=1.6 \times 10^{-5}$, $R^2=0.9989$), respectively. We can see that despite complicated spectra with asymmetric peaks and varying features, the fits are extremely good. 

\section{Machine Learning Framework}
Following the procedure in Ref. \cite{mahani2023data}, we employ a ML framework to predict the reflectivity of surface Bragg gratings based on design parameters using simulation data. The workflow involves splitting the dataset into training and testing subsets, where the training set (90\% of total number of data points) is used to train the ML model and the testing set (the remaining 10\%  of the total number of data points) is used to evaluate its performance. The Optimized XGBoost ML model (high performing algorithm on small data \cite{mahani2023data}) was trained using three input parameters, groove depth, groove width, and order number of the gratings to predict the SVD fit parameters and the corresponding reflectance spectra.
\begin{figure}[htbp]
\centering
\includegraphics[height=5.8in]{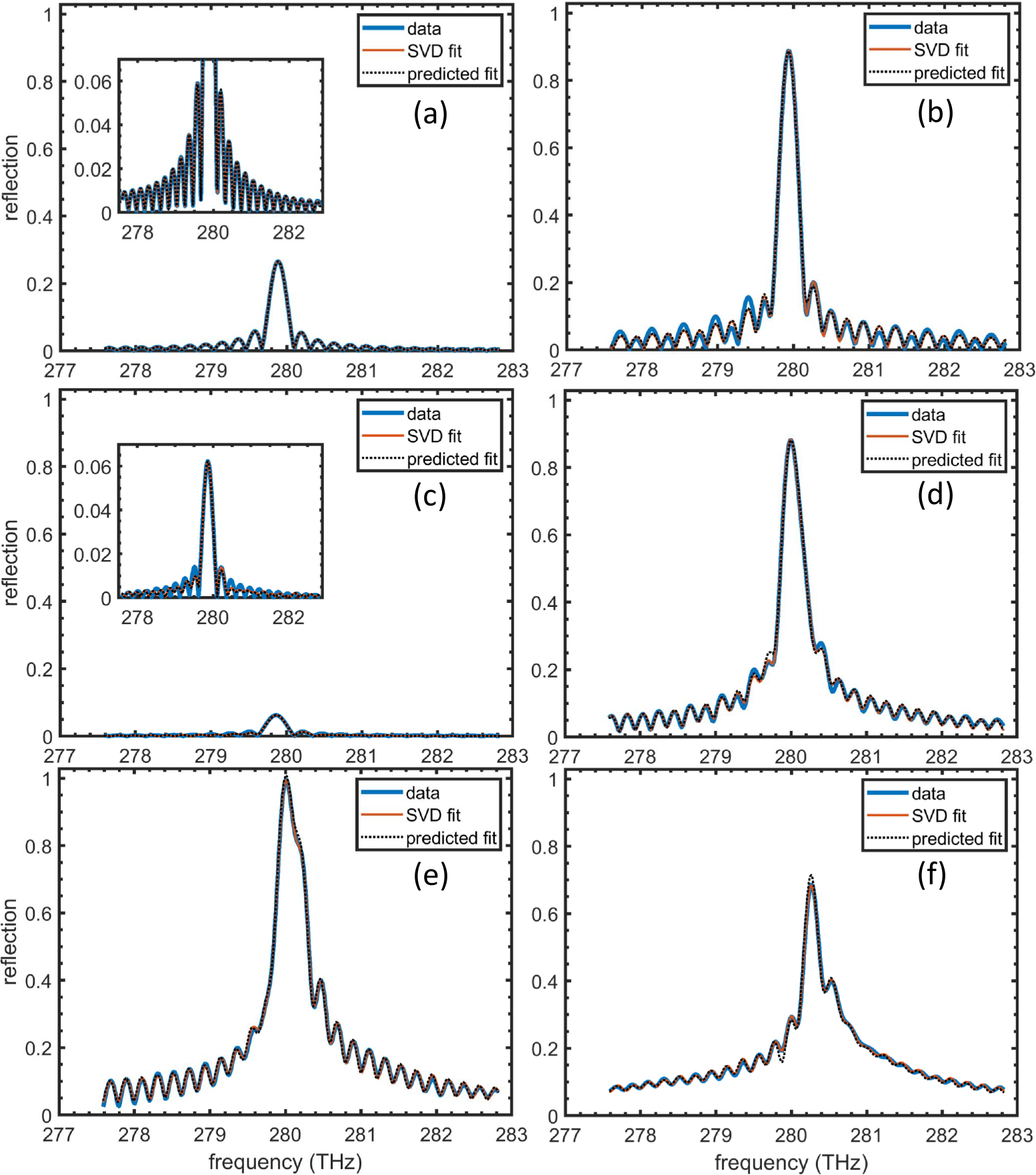}
\caption{Prediction of reflectance spectra using optimized XGBoost ML. Panels (a)-(f) are similar to panels in figure \ref{fig:ref_pred}, with ML predictions added (refer to in the legend as predicted fit). ML prediction metrics for a random test set (10\% of total data points) that includes panels (a)-(f) are $\textnormal{MSE}=2.95 \times 10^{-6}$ and $R^2=0.983$.}
\label{fig:ref_pred2}
\end{figure}

XGBoost, initially introduced by Chen and Guestrin \cite{chen2016xgboost}, is a highly efficient and flexible gradient-boosting algorithm. It implements a distributed gradient-boosted decision tree (GBDT) model, which is an ensemble learning method that combines multiple decision trees (specifically regression trees) to enhance predictive accuracy. Unlike the random forest model \cite{breiman2001random}, where trees are built independently, GBDTs construct trees in a sequential manner, with each new tree learning from the errors of the previous ones. In our work, we fine-tuned the XGBoost parameters using GridSearchCV, a Scikit-learn function \cite{pedregosa2011scikit}, which systematically searches for the optimal parameter values within a predefined grid. The performance of this model was assessed using standard metrics, namely, MSE and R\textsuperscript{2}.

After training was done on 90\% of the data, we test the model on the remaining 10\%. Since both training and test sets are chosen randomly, we repeat this procedure fifty times and then take the average. We obtain $R^2=0.984$ with a standard deviation of $\textnormal{std.}=0.003$ and $\textnormal{MSE}=2.77 \times 10^{-6}$ with a standard deviation of $std.=5.84 \times 10^{-7}$ for prediction of the fit parameters. These metrics indicate that the model predicts with high accuracy and captures the variations in the data very well. The results of the predictions are plotted in Fig.~\ref{fig:ref_pred2}. This figure visually shows the predicted fits agree very well with the SVD fit. The result demonstrates the overall reliability of the SVD method combined with the Optimized XGBoost for predicting complex optical spectra with a high precision.
\section{Selection of Data Points using Bayesian Approach}
\label{bayesian}
In our prior work \cite{mahani2024optimizing}, we investigated the limitations of traditional data acquisition methods, such as uniform and random sampling, for acquiring data used for training ML models in high-dimensional parameter spaces. These methods often require extensive data points, leading to resource-intensive simulations and experiments, particularly in fields where data collection is challenging. To address this, we employed a Bayesian-based approach to optimize data acquisition. This method leverages Bayesian Optimization (BO) to iteratively identify the most informative data points, drastically reducing the number of simulations required while maintaining the predictive accuracy of the ML models.

The database chosen using BO, Bayesian-based dataset (BBD), is constructed in a way that is minimal yet highly informative for training ML models. Unlike traditional methods that either uniformly distribute or randomly sample data points, the BBD is built using Gaussian process regression (GPR) to model the underlying relationship between input parameters and output responses. BO then selects new data points with maximum uncertainty, thereby ensuring that each new simulation adds new information to the model's training process. This method has shown that ML models trained on BBDs can achieve equal performance with much fewer data points compared to those trained on uniformly or randomly sampled databases \cite{mahani2024optimizing}.

We apply BO to our database (contains 2736 data points generated uniformly for each input parameters) to extract minimum number of data points, where ML can reach the accuracy of $R^2=0.9$. This value of accuracy can be adjusted to any required value. Our simulations show that if the data points chosen using BO, we can train the ML model with only 600 data points instead of 950 data points with random mesh to get the same performance metric ($R^2=0.9$), showing 37\% decrease in the number of data required.

\begin{figure}[htbp]
\centering
\subfloat[]{\includegraphics[height=1.7in]{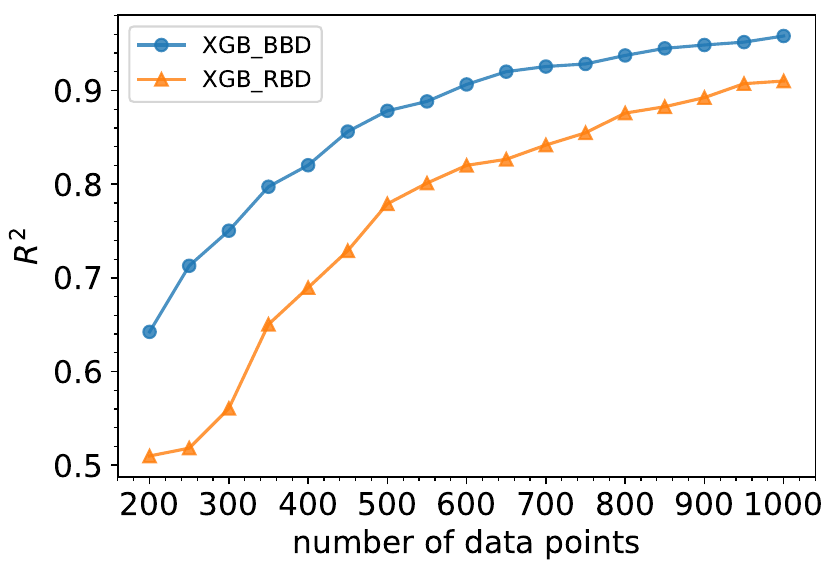}}
\subfloat[]{\includegraphics[height=1.7in]{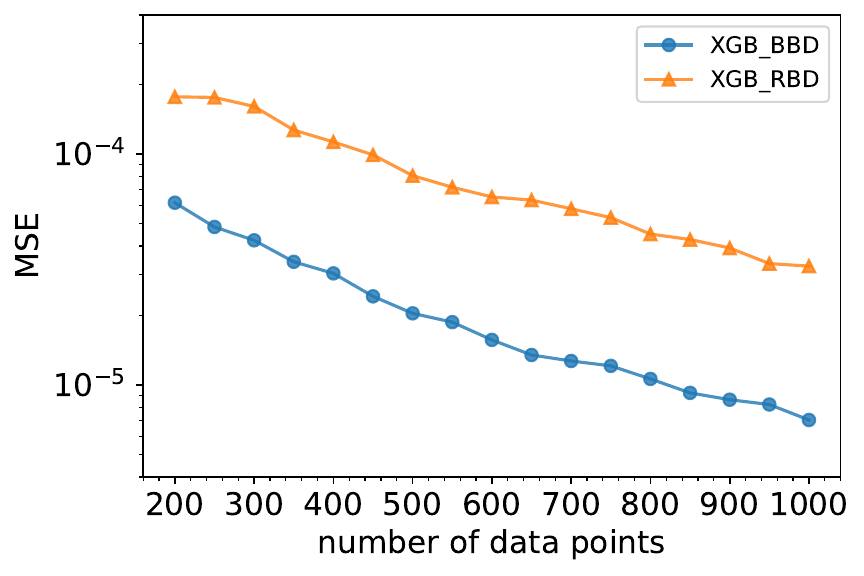}}
\caption{Performance comparison between XGBoost ML model trained on Bayesian based database (BBD) vs.\ random based database (RBD). The number of data points have been increased up to 1000 data points. (a) and (b) $R^2$ and MSE as a function of number of data points, respectively. To achieve an accuracy of $R^2=0.9$, the number of data points obtained with RBD is 950, while this number for BBD is 600 data points, showing 37\% decrease in the number of data required.}
\label{fig:BO_performance}
\end{figure}

Figure \ref{fig:BO_performance}(a)-(b) show the performance comparison ($R^2$ and MSE) between XGBoost ML model trained on BBD versus random based database (RBD). To achieve an accuracy of $R^2=0.9$, the ML model needs 37\% fewer data points for training, if they are chosen using BBD approach compared with RBD. This performance difference is more visible when values of MSE are compared.
In applying this method, we demonstrated that the BBD approach reduces the required data for ML training without loosing the predictive accuracy of ML model. This is particularly crucial for applications requiring high precision, such as the design of optical devices where the predictive capability of the ML model directly impacts the efficiency of the design process. 

\begin{figure}[htbp]
\centering
\subfloat[]{\includegraphics[height=3.6 in]{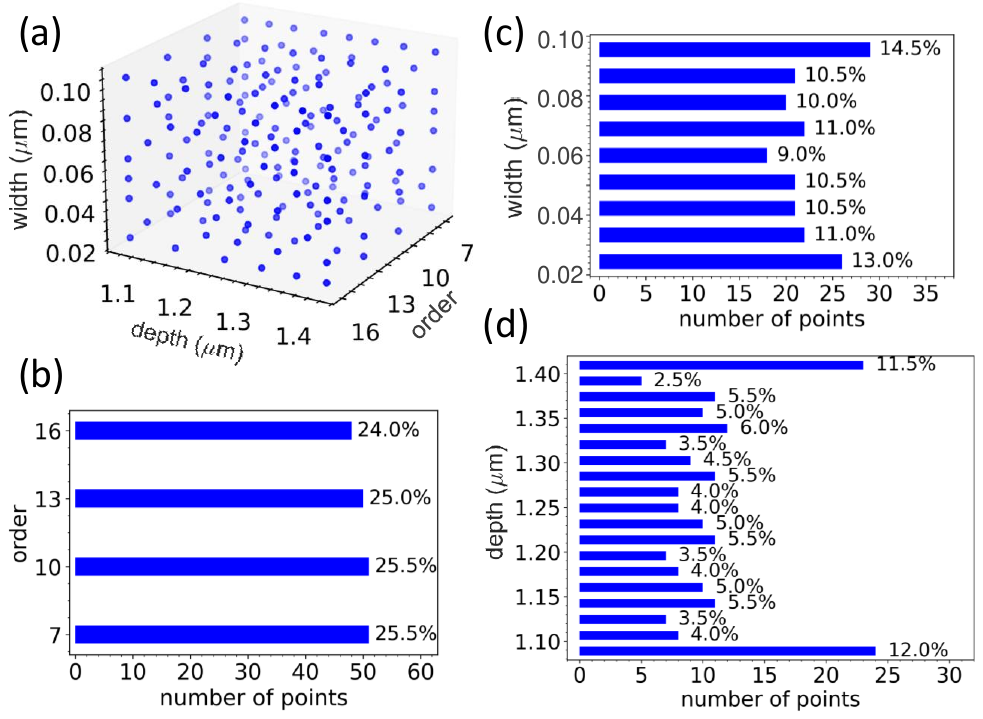}}
\caption{Statistical analyses of the first 200 data points chosen by Bayesian approach. (a) the Bayesian distribution of the first 200 data points showing where in the design space, the Bayesian approach selects the data. (b) the distribution of the data categorized by different order number, showing the percentage of data points associated with each order. (c) the distribution of data points for various width interval. (d) the distribution of data points for various depth values.}
\label{fig:BBD_stats}
\end{figure}
In Figure \ref{fig:BBD_stats}(a), the Bayesian distribution of the first 200 data points is presented. This figure shows the probabilistic distribution of the selected points, allowing us to assess where in the design space, the Bayesian approach selects the data. 
In Figures \ref{fig:BBD_stats}(b), \ref{fig:BBD_stats}(c), and \ref{fig:BBD_stats}(d), we show the distribution of points based on specific geometrical parameters. Figure \ref{fig:BBD_stats}(b) illustrates the distribution of the data categorized by different order number, showing the percentage of data points associated with each order. We see that BO selects almost equally from each order number. Figure \ref{fig:BBD_stats}(c) presents the distribution of data points for various width interval, showing that data points are more concentrated at the edges of the defined intervals. Similarly, Figure \ref{fig:BBD_stats}(d) demonstrates the distribution of data points for various depth values. This panel also exhibits a higher distribution near the minimum and maximum bounds. This indicates that the BO tends to select data points at these extremes. The physical significance of this behavior lies in the fact that both the minimum and maximum depth values have a larger effect on the spectral response of the system, inducing the strongest variations. Depth, in particular, strongly influences the spectra, with extreme values producing distinct spectral signatures. Consequently, these edge points in depth are critical in the analysis and contribute substantially to the variation observed in the dataset.

\section{Conclusion}
In this study, we generated a database for the design of Bragg gratings with varying groove depths, widths, and order numbers, resulting in 2736 data points. Each data point corresponds to a unique set of parameters and its associated reflectance spectrum. To efficiently parametrize the entire reflectance spectra, we applied a truncated SVD with $K=7$ modes leading to $14$ parameters. This allowed us to reduce the dimensionality of the data while retaining the most significant features of the reflectance spectra, thus facilitating a more efficient and accurate analysis of the optical properties of the Bragg gratings.

Subsequently, we employed the Bayesian selection approach to optimize the data acquisition process by selecting a limited number of data points from the full dataset. This method strategically chooses data points with the highest information content, thereby maximizing the predictive power of the model while minimizing the number of required simulations. 

Finally, we applied the XGBoost ML model to the limited dataset selected through BO. XGBoost was chosen for its good performance with scarce data and its ability to model complex relationships between the input parameters and the reflectance spectra. By training on the reduced dataset, XGBoost effectively predicted the full reflectance spectra across the entire parameter space with high accuracy. 

The combination of SVD for dimensionality reduction, Bayesian optimization for data selection, and XGBoost for prediction provides an efficient and accurate framework for the design of Bragg gratings, demonstrating the potential of this integrated approach for solving complex optical design problems.\\

\textbf{Author contributions:}
MRM, IN, TF and AW conceived and planned the research. MRM carried out the method development, optimizations and ML simulations. MRM and TF carried out the SVD fit analyses. IN performed the FDTD Bragg grating simulations and generated the database. All authors contributed to the interpretation of the results, manuscript writing, and provided feedback.\\

\textbf{Research funding:} This work was supported by the VDI Technologiezentrum GmbH with funds provided by the Federal Ministry of Education and Research under grant no. 13N14906 and the DLR Space Administration with funds provided by the Federal Ministry for Economic Affairs and Climate Action (BMWK) under Grant No 50WK2272.\\

\textbf{Conflict of interest statement:} The authors declare no conflicts of interest regarding this article.\\

\bibliography{References}

\end{document}